\documentclass[aps,twocolumn,showpacs,prb,superscriptaddress,floatfix,longbibliography]{revtex4-2}

\usepackage{longtable}
\usepackage{multirow}
\usepackage{enumitem}
\usepackage{amsthm}
\usepackage{amsmath}
\usepackage{amsfonts}
\usepackage[ruled,vlined]{algorithm2e}
\usepackage{comment}
\usepackage{hyperref}
\usepackage{tabularx}
\usepackage{float}
\usepackage{xcolor}
\usepackage{tikz}
\usetikzlibrary{shapes.geometric, arrows}
\usepackage[normalem]{ulem}

\newtheorem{thrm}{Theorem}

\usepackage{graphicx}
\usepackage{dcolumn}
\usepackage{bm}
\usepackage{bbm}
\usepackage{hyperref}

\begin{document}

\author{Lexin Ding}
\affiliation{Faculty of Physics, Arnold Sommerfeld Centre for Theoretical Physics (ASC),\\Ludwig-Maximilians-Universit{\"a}t M{\"u}nchen, Theresienstr.~37, 80333 M{\"u}nchen, Germany}
\affiliation{Munich Center for Quantum Science and Technology (MCQST), Schellingstrasse 4, 80799 M{\"u}nchen, Germany}

\author{Stefan Knecht}
\affiliation{Algorithmiq Ltd., Kanavakatu 3C, FI-00160 Helsinki, Finland}
\affiliation{ETH Z{\"u}rich, Laboratory for Physical Chemistry, Vladimir-Prelog-Weg~2, 8093 Z{\"u}rich, Switzerland}

\author{Christian Schilling}
\email{c.schilling@lmu.de}
\affiliation{Faculty of Physics, Arnold Sommerfeld Centre for Theoretical Physics (ASC),\\Ludwig-Maximilians-Universit{\"a}t M{\"u}nchen, Theresienstr.~37, 80333 M{\"u}nchen, Germany}
\affiliation{Munich Center for Quantum Science and Technology (MCQST), Schellingstrasse 4, 80799 M{\"u}nchen, Germany}

\begin{abstract}
We propose an effective quantum information-assisted complete active space optimization scheme (QICAS). What sets QICAS apart from other correlation-based selection schemes is (i) the use of unique measures from quantum information that assess the correlation in electronic structures in an unambiguous and predictive manner, and (ii) an orbital optimization step that minimizes the correlation discarded by the active space approximation. Equipped with these features QICAS yields for smaller correlated molecules sets of optimized orbitals with respect to which the CASCI energy reaches the corresponding CASSCF energy within chemical accuracy. For more challenging systems such as the Chromium dimer, QICAS offers an excellent starting point for CASSCF by greatly reducing the number of iterations required for numerical convergence. Accordingly, our study validates a profound empirical conjecture:  the energetically optimal non-active spaces are predominantly those that contain the least entanglement.
\end{abstract}

\title{Quantum Information-Assisted Complete Active Space Optimization (QICAS)}

\maketitle

Describing the electronic structure of strongly correlated molecular systems is one of the major challenges in modern-day quantum chemistry (QC). In contrast to weakly correlated systems, the ground state of a strongly correlated system can no longer be accurately represented by a single reference configuration, such as the one obtained by solving the Hartree-Fock (HF) equations. Instead, the interaction between different configurations must be taken into account, commonly referred to as \textit{static correlation}. Although a full configuration interaction (full CI, or FCI) approach is not realistic for systems consisting of more than a couple of dozens of orbitals at half-filling\cite{vogiatzis2017pushing}, it is possible to recover to a large degree the electron correlation by restricting the interacting configurations to those that only differ in the occupations of a fraction of orbitals. This amounts to the so-called complete active space configuration interaction (CASCI) approach. If, in addition, the underlying molecular orbital basis is optimized in a self-consistent fashion, CASCI turns into the variationally superior complete active space self-consistent field (CASSCF) approach\cite{roos1980complete,werner1985second,siegbahn1981complete,malmqvist1989casscf,kreplin2019second,kreplin2020mcscf}.

That being said, determining accurate active spaces in practice is a crucial challenge since the quality as well as rate of convergence of a CASCI/CASSCF calculation are highly sensitive to the choice of (initial) active orbitals\cite{review21}. Moreover, the purpose of choosing an active space is to recover correlation effects sufficiently well such that the system at hand is described adequately by a CAS model. To this end, chemical knowledge of the individual systems at hand is often required\cite{roos1987complete,pierloot2003caspt2,veryazov2011select,knizia17} for active space selection. Yet, such \textit{a priori} knowledge of the system may not always be available especially when the system is large, and the lack thereof severely hinders the black-box applications of CAS methods.

Nonetheless, there is hope that the electronic structure estimated by post-HF solutions can be exploited to remove this roadblock: Based on a correlated wavefunction, active orbitals may be selected by means of orbital ``observables" (e.g.~their occupation numbers) that provide an estimate of orbital correlation effects\cite{jensen1988second,pulay1988uhf,pulay89,abrams2004natural,pulay15,zou2020automatic}.
Consequently, diagnostic tools offered by quantum information (QI) which quantify the orbital correlations in a concise manner \cite{legeza03,reiher12entanglement,reiher16,boguslawski2017multi,ding2020correlation,ding2020concept,ding2022quantum,bensberg2023corresponding} may hold the key to effective black-box active space selection protocols.
One prime example of such tools is the von Neumann entropy
\begin{equation}
    \begin{split}
    S(\rho_i) &= -\mathrm{\rho_i\log(\rho_i)},
    \\
    \rho_i &= \mathrm{Tr}_{\backslash\{\phi_i\}}[|\Psi_0\rangle\langle\Psi_0|] \label{eqn:entropy}
    \end{split}
\end{equation}
of the reduced density matrix $\rho_i$ for orbital $\phi_i$, obtained by tracing out all other orbital degrees of freedom of the ground state $|\Psi_0\rangle$.
This quantity is precisely the entanglement between orbital $\phi_i$ and the complementary system consisting of all other orbitals.
For a more detailed account of the concepts of fermionic reduced density matrices and orbital entanglement,
we refer the readers to Refs.~\cite{friis2016reasonable,boguslawski2015orbital,ding2020concept}.

There is a caveat, though, as extracting orbital correlation inevitably requires a multireference description of the system. Such a description obviously should not be an exact, but rather an affordable approximate one (otherwise the purpose of the general idea would be defeated). The latter is nowadays accessible at reasonable computational cost, thanks to the advent of the matrix product state (MPS) ansatz\cite{cirac07mps,schollwock2011density} underlying the density matrix renormalization group (DMRG) approach\cite{white92density,schollwock05review} in QC\cite{white1999ab,marti2010density,chan2011density,wouters2014density,baia20a}.

To provide first evidence for the promising prospects of this QI-inspired paradigm,
Stein and Reiher determined in \cite{reiher16} suitable active space sizes based on approximate ground states $|\Psi_0\rangle$ obtained from preceding DMRG calculations with low bond dimensions.
To be more specific, it was their crucial observation that for a given system with reference orbitals $\{\phi_i\}$ the corresponding orbital entropy profile $\{S(\rho_i)\}$
reveals a plateau structure and thus determines reasonable choices for the active space size.

Despite the fruitful application of the single orbital entropy \eqref{eqn:entropy} in the context of \emph{fixed} orbital reference bases, the full analytic potential of QI quantities such as $S(\rho_i)$\ is yet to be exploited for active space \textit{optimization}. In particular, it is still an open  challenge to reveal and establish a link that relates a correlation-based measure of the ``goodness''/``badness'' of an active space in a predictable manner to the accompanying CASCI energy.
Without such a measure, the which the active orbitals are selected, cannot be systematically \textit{optimized}, and therefore the ``perfect'' active space will always be out of reach.

In our letter, we provide this pivotal missing link by proposing a tailored measure that evaluates the quality of a given active space based on orbital entanglement entropies. This in turn then allows us to propose a quantum information-assisted complete active space optimization (QICAS) method with the following appealing features: QICAS (1) requires little system-dependent knowledge, (2) chooses active orbitals not from a \textit{fixed} but a \textit{variable} set of reference orbitals which is \textit{optimized} based on our QI-motivated cost function, and (3) produces active spaces starting from which a subsequent CASSCF calculation is either practically converged or requires much fewer iterations. Accordingly, our work offers a change of paradigm in addressing the ground state problem: QICAS could be added as an intermediate layer between initial HF computation and final post HF-treatment in order to improve the overall accuracy and efficiency.

We start by considering an $N$-electron ground state problem characterized by the electronic Hamiltonian $\hat{H}$ defined in second quantization with respect to an \textit{ordered} basis $\mathcal{B}$ of $D$ molecular orbitals $\mathcal{B}=\{\phi_i\}_{i=1}^D$. A complete active space (CAS) problem is then uniquely determined by the tuple $(N_\mathrm{CAS}, D_\mathrm{CAS})$ of number $N_\mathrm{CAS}$ of active electrons and number $D_\mathrm{CAS}$ of active orbitals, together with $\mathcal{B}$. Based on these parameters, it is assumed that the first $(N\!-\!N_\mathrm{CAS})/2$ orbitals in $\mathcal{B}$ are fully occupied (closed), whereas the last $D\!-\!D_\mathrm{CAS}\!-\!(N\!-\!N_\mathrm{CAS})/2$ orbitals are empty (virtual). That is, the correlation effect involving the closed and virtual orbitals is ignored, and only that involving the $D_\mathrm{CAS}$ active orbitals is accounted for by a full CI calculation within the active space. For a basis $\mathcal{B}$, we denote this method as $\mathcal{B}$-CASCI($N_\mathrm{CAS}$, $D_\mathrm{CAS}$). Using the orbital reduced states $\rho_i$ of the full CI ground state $|\Psi_0\rangle$, the correlation contribution involving each orbital $\phi_i$ and the rest of the system can be exactly quantified as the corresponding single orbital entropy $S(\rho_i)$ (see Refs.~\cite{legeza03,rissler06orbital,reiher12entanglement,boguslawski2015orbital,ding2020concept}). From this, we define the out-of-CAS correlation $F_\mathrm{QI}(\mathcal{B})$ as the sum of orbital entropies over all non-active (closed and virtual) orbitals
\begin{equation}
    F_\mathrm{QI}(\mathcal{B}) \equiv \sum_{i \in \mathcal{N}} S(\rho_i), \label{eqn:cost_QICAS}
\end{equation}
where $\mathcal{N}$ conveniently denotes the set of indices of the non-active orbitals.
The quantum information meaning of Eq.~\eqref{eqn:cost_QICAS} is more complicated than just orbital entanglement,
as it is the hybrid sum of the multipartite correlation\cite{watanabe1960information} within the non-active space,
\begin{equation}
\begin{split}
    I_{\mathcal{N}}(\mathcal{B}) &= \sum_{i \in \mathcal{N}} S(\rho_i) - S(\rho_{\mathcal{N}}),
    \\
    \rho_{\mathcal{N}} &= \mathrm{Tr}_{\mathcal{A}}[|\Psi_0\rangle\langle\Psi_0|],
\end{split}
\end{equation}
and the entanglement between the active $\mathcal{A}$ and non-active space $\mathcal{N}$
\begin{equation}
    E_{\mathcal{A}|\mathcal{N}}(\mathcal{B}) = S(\rho_\mathcal{N}).
\end{equation}
These quantities are nothing else than the orbital correlation and entanglement entirely discarded by the $\mathcal{B}$-CASCI($N_\mathrm{CAS}$,\:$D_\mathrm{CAS}$) scheme.
In the following we shall refer to $F_\mathrm{QI}(\mathcal{B})$ as the out-of-CAS orbital correlation for simplicity.

In different bases $\mathcal{B}$, the same CAS($N_\mathrm{CAS}$,\:$D_\mathrm{CAS}$) scheme discards different values of out-of-CAS orbital correlation $F_\mathrm{QI}(\mathcal{B})$.
Likewise, the $\mathcal{B}$-CASCI($N_\mathrm{CAS}$,\:$D_\mathrm{CAS}$) energy varies significantly with the ordered basis $\mathcal{B}$.
Therefore, in order to facilitate the optimization of the active space, a crucial step of our work is to establish a quantitative connection between the CASCI energy $E_0^\mathrm{CASCI}(\mathcal{B})$ and the out-of-CAS correlation $F_\mathrm{QI}(\mathcal{B})$.
Intuitively, one would expect a CAS scheme discarding less correlation should lead to a lower CASCI energy. In particular, if the CASCI energy $E^{\mathrm{CASCI}}_0(\mathcal{B})$ and the full CI energy $E_0^{\mathrm{FCI}}$ were identical (in some hypothetical system), i.e, the CAS ansatz was exact, then $S(\rho_i) = 0$ would follow for all $i\in\mathcal{N}$ in the full CI ground state $|\Psi_0\rangle$.
To see this, notice that $|\Psi_0\rangle$ would be a CAS ansatz state with definite occupation numbers on the non-active orbitals. Therefore the reduced states $\rho_i$ for $i \in \mathcal{N}$ are pure states (either $|0\rangle\langle 0|$ for virtual orbitals or ${|\!\uparrow\downarrow\rangle\langle\uparrow\downarrow\!|}$ for closed orbitals), leading to $S(\rho_i)=0$. Conversely, when the discarded correlation does not vanish, the energy difference $\Delta E(\mathcal{B}) \equiv E^{\mathrm{CASCI}}_0(\mathcal{B}) - E_0^{\mathrm{FCI}}$ must also be finite. The energy difference $\Delta E(\mathcal{B})$ is then the remaining part of the correlation energy that the active space fails to capture. One of our key discoveries is that $\Delta E(\mathcal{B})$ is bounded from above by $F_\mathrm{QI}(\mathcal{B})$. This pivotal result is formulated rigorously in the following theorem, which lies at the heart of our entire work.


\begin{thrm} \label{thrm}
For any molecular system with electronic Hamiltonian $\hat{H}$ the following energy-correlation relation
\begin{equation}
\Delta E(\mathcal{B}) \equiv E_0^\mathrm{CASCI}(\mathcal{B}) - E_0^\mathrm{FCI} \leq k(\hat{H}) F_\mathrm{QI}(\mathcal{B})
\end{equation}
holds for all orbital bases $\mathcal{B}$, where $k(\hat{H})$ is a constant independent of $\mathcal{B}$ and $F_\mathrm{QI}(\mathcal{B}) \equiv \sum_{i \in \mathcal{N}} S(\rho_i)$.
\end{thrm}

The proof of Theorem \ref{thrm} is reserved to Appendix \ref{app:proof} due to its technical character. The key message of this theorem is that the relation between $F_\mathrm{QI}(\mathcal{B})$ and $E^{\mathrm{CASCI}}_0(\mathcal{B})$ is monotonic in the following sense: as the discarded correlation $F_\mathrm{QI}(\mathcal{B})$ decreases by varying $\mathcal{B}$, the upper bound of the corresponding CASCI energy $E^{\mathrm{CASCI}}_0(\mathcal{B})$ also decreases. Theorem \ref{thrm} therefore unequivocally justifies $F_\mathrm{QI}(\mathcal{B})$ as an alternative cost function for active/non-active orbital optimization.

\begin{figure}[t]
    \centering
    \includegraphics[scale=0.3]{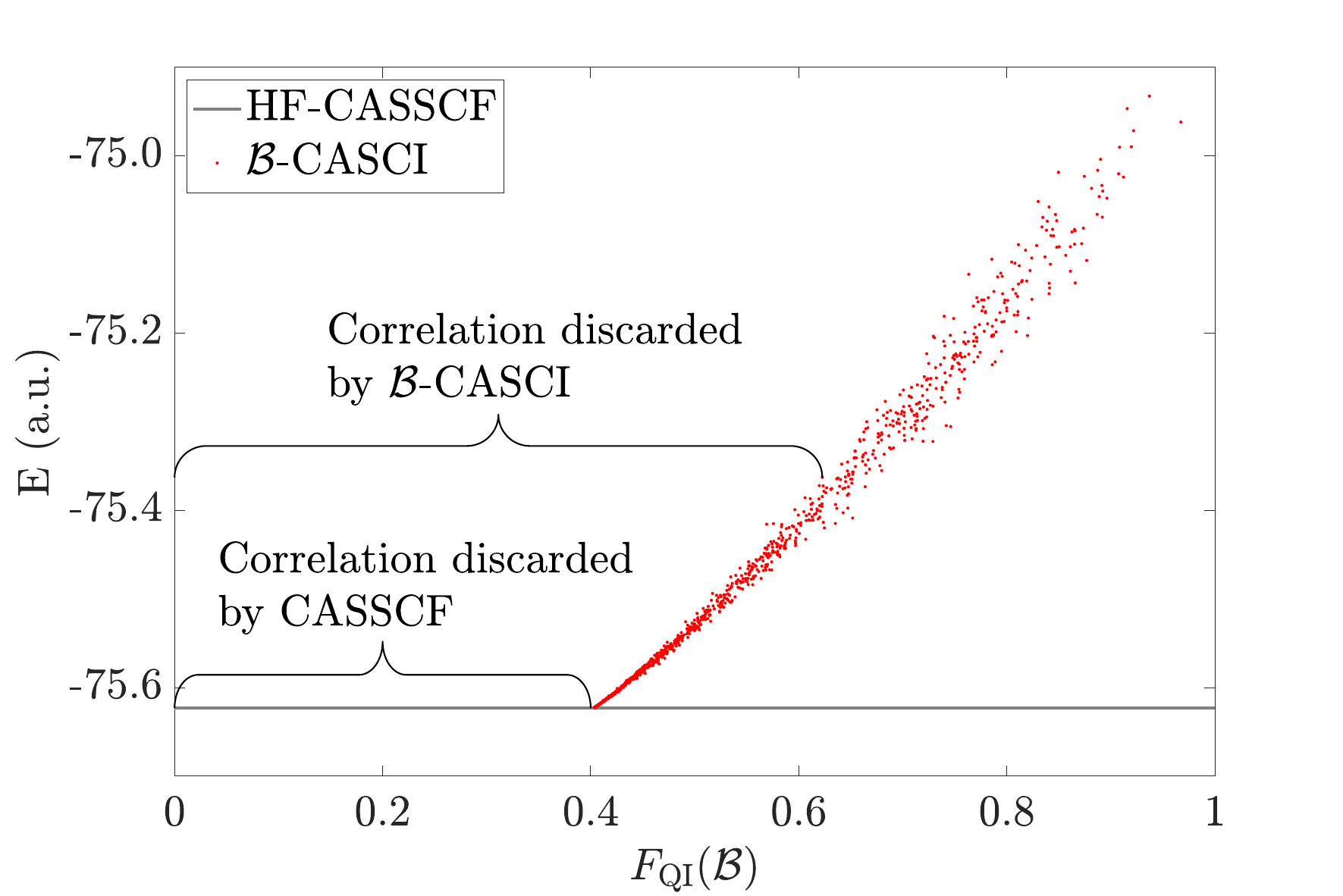}
    \caption{$\mathcal{B}$-CASCI(8,8) energy for $\mathrm{C_2}$ at $\mathrm{R} = 1.243\,\mathrm{\AA}$ in the cc-pVDZ basis set versus out-of-CAS correlation $F_\mathrm{QI}(\mathcal{B})$. 1000 random orbital bases $\mathcal{B}$ are sampled. The clear linear behaviour confirms the significant relevance of Theorem \ref{thrm}.}
    \label{fig:qicas_c2vdz_r1.243_perturb}
\end{figure}

\begin{figure*}[htb]
    \centering
    \includegraphics[scale=0.35]{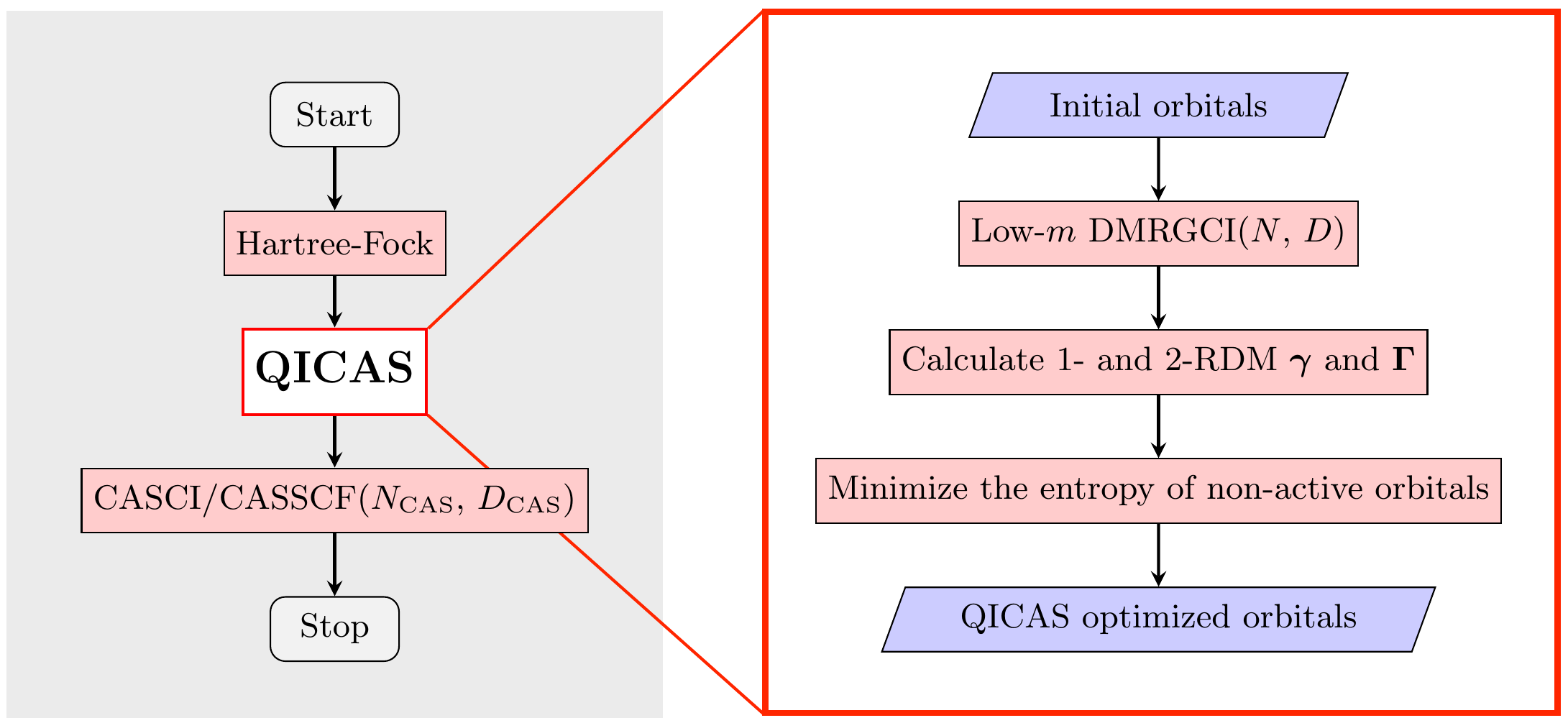}
    \caption{Flowchart of the QICAS subroutine and its applications as a post-HF treatment to prepare orbitals for CASCI/CASSCF calculations.
    The DMRG step is performed on $D$ orbitals and $N$ electrons, with an MPS ansatz of bond dimension $m$.}
    \label{fig:schematics}
\end{figure*}

To showcase this important result,
we demonstrate in the following this monotonicity with the $\mathrm{C_2}$ molecule at equilibrium geometry ($\mathrm{R}=1.243\,\mathrm{\AA}$\cite{huber2013molecular}) in the cc-pVDZ basis set\cite{dunning1989gaussian} and a CAS(8,\:8) scheme.
In Figure \ref{fig:qicas_c2vdz_r1.243_perturb},
we present the out-of-CAS correlation $F_\mathrm{QI}(\mathcal{B})$ of 1000 randomly sampled orbital bases $\mathcal{B}$ around the minimizer $\mathcal{B}^\ast$ of $F_\mathrm{QI}(\mathcal{B})$, against the corresponding CASCI energy $E_0^\mathrm{CASCI}(\mathcal{B})$.
The $x$-value of each data point can be interpreted as the orbital correlation discarded by the CASCI solution in this basis.
The overall trend is apparent: a lower $F_\mathrm{QI}(\mathcal{B})$ generally leads to a lower CASCI energy. The relation is roughly linear with some dispersion when $F_\mathrm{QI}(\mathcal{B})$ is large, but in the region $F_\mathrm{QI}(\mathcal{B}) < 1/2$ the dispersion is much smaller and the linearity is much better. Even more remarkably, when the corresponding value of $F_\mathrm{QI}(\mathcal{B})$ is at the minimum (around 0.4), the CASCI energy matches the CASSCF energy. This rationalizes a profound empirical expectation: The energetically optimal non-active spaces are predominantly those that contain the least entanglement.




In light of Theorem 1 and its crucial predictions, we propose and establish in the following an effective quantum information-assisted complete active space optimization scheme (QICAS). The steps constituting QICAS are summarized as a flowchart in Figure \ref{fig:schematics}.  It starts with first and foremost a low bond dimension MPS approximation $|\Psi_0\rangle$ to the ground state obtained via QC-DMRG based on an initial orbital reference basis $\mathcal{B}_0$ (e.g., the HF basis $\mathcal{B}_{\mathrm{HF}}$). To this end, we use the block2\cite{zhai2021low} module integrated within the \textsc{PySCF} package\cite{sun2018pyscf}. The accuracy and computational cost can be systematically tuned by varying the bond dimension $m$ of the MPS. The state $|\Psi_0\rangle$ then captures most of the static correlation effects in the basis $\mathcal{B}_0$, which are encoded in the corresponding 1- and 2-particle reduced density matrix $\bm{\gamma}$ and $\bm{\Gamma}$ as well.
It is then the central step of QICAS to reorganize the orbital correlation structure in order to determine a more favorable separation into active and non-active space. This is achieved by minimizing the out-of-CAS correlation $F_\mathrm{QI}(\mathcal{B})$ in Eq.~\eqref{eqn:cost_QICAS} using orbital rotations. The set of resulting optimal orbitals is denoted as $\mathcal{B}^\ast$ and a non-active orbital in $\mathcal{B}^\ast$ is classified as closed (virtual) if its occupancy is larger (smaller) than 1.



The huge potential of QICAS lies in the simplicity of the cost function $F_\mathrm{QI}(\mathcal{B})$. Due to the particle number and spin symmetry of the electronic Hamiltonian, any single orbital reduced state $\rho_i$ is already diagonal in the ``local'' eigenbasis $\{|\,0\rangle_i, {|\!\uparrow\rangle_i}, {|\!\downarrow\rangle_i}, |\!\uparrow\downarrow\rangle_i\}$, and the respective eigenvalues $(\lambda_k^{(i)})_{k=0}^3$ are given by\cite{boguslawski2015orbital,ding2020concept}
\begin{equation}
    \begin{split}
        \rho_i \,|0\rangle_i &= (1-\gamma^{i}_i-\gamma^{\bar{i}}_{\bar{i}} + \Gamma^{i\bar{i}}_{i\bar{i}}) \,|0\rangle_i \equiv \lambda^{(i)}_0 |0\rangle_i,
        \\
        \rho_i \,|\!\uparrow\rangle_i &= (\gamma^{i}_i-\Gamma^{i\bar{i}}_{i\bar{i}}) \,|\!\uparrow\rangle_i \equiv \lambda^{(i)}_{1}|\!\uparrow\rangle_i,
        \\
        \rho_i \,|\!\downarrow\rangle_i &= (\gamma^{\bar{i}}_{\bar{i}} - \Gamma^{i\bar{i}}_{i\bar{i}}) \,|\!\downarrow\rangle_i \equiv \lambda^{(i)}_2 |\!\downarrow\rangle_i,
        \\
        \rho_i \,|\!\uparrow\downarrow\rangle_i &= \Gamma^{i\bar{i}}_{i\bar{i}} \,|\!\uparrow\downarrow\rangle_i \equiv \lambda^{(i)}_{3}|\!\uparrow\downarrow\rangle_i.
    \end{split}\label{eqn:ordms}
\end{equation}
Here, $i$ ($\bar{i}$) denotes the spin-orbital labeled by the spatial orbital index $i$ and spin $\uparrow$ ($\downarrow$), and the entries of $\bm{\gamma}$ and $\bm{\Gamma}$ in the basis $\mathcal{B}$ are defined as
\begin{equation}
    \begin{split}
        \gamma^{i}_{j} &= \langle \Psi_0 | f^{\dagger}_i f^{\phantom{\dagger}}_j |\Psi_0\rangle,
        \\
        \Gamma^{ij}_{kl} &= \langle \Psi_0 | f^{\dagger}_i f^{\dagger}_j f^{\phantom{\dagger}}_l f^{\phantom{\dagger}}_k |\Psi_0\rangle, \label{eqn:prdms}
    \end{split}
\end{equation}
where $f^{(\dagger)}_{i/\bar{i}}$ annihilates (creates) an electron in the $i$$\uparrow$/$i$$\downarrow$ spin-orbital. Accordingly, the cost function $F_\mathrm{QI}(\mathcal{B})$ can be expressed as
\begin{equation}
    F_\mathrm{QI}(\mathcal{B}) = - \sum_{i\in \mathcal{N}} \sum_{k=0}^3 \lambda^{(i)}_k \log(\lambda^{(i)}_k). \label{eqn:cost_explicit}
\end{equation}
Some crucial remarks are in order here. First, the $\mathcal{B}$-dependence of the right-hand side of \eqref{eqn:cost_explicit} is implicitly encoded in the eigenvalues of the orbital reduced density matrices $\rho_i$. Second, $F_\mathrm{QI}(\mathcal{B})$ can easily be computed for \textit{any} $\mathcal{B}$. One just needs to rotate $\bm{\gamma}$ and $\bm{\Gamma}$ from the original reference basis $\mathcal{B}_0$ to $\mathcal{B}$ and then extract the relevant entries for Eq.~\eqref{eqn:ordms}.
Third, the computation and optimization of $F_\mathrm{QI}(\mathcal{B})$ does \emph{not} invoke the Hamiltonian. Instead, the information of the latter is implicitly encoded in the approximated ground state $|\Psi_0\rangle$, which is used only once before the optimization for computing $\bm{\gamma}$ and $\bm{\Gamma}$ in the basis $\mathcal{B}_0$. All these features highlight the huge potential of QICAS for optimizing active spaces more efficiently than CASSCF, as in the latter the energy and its gradients need to be computed many times.

In the following, we propose and test two applications of QICAS as post-HF treatments (see Figure \ref{fig:schematics}):
\begin{enumerate}[label=(\textbf{\Alph*})]
    \item QICAS followed by a CASCI calculation (HF-QICAS-CASCI).

    \item QICAS performed on a subset $\mathcal{S}$ of orbitals (${D_\mathrm{CAS}<|\mathcal{S}|\leq D}$) to prepare a potentially more advantageous starting point for CASSCF (HF-QICAS-CASSCF),
    where $|\mathcal{S}|$ should be several times larger than $D_{\mathrm{CAS}}$ to ensure a reasonable performance.
\end{enumerate}

\begin{figure*}[htb]
    \centering
    \includegraphics[scale=0.31]{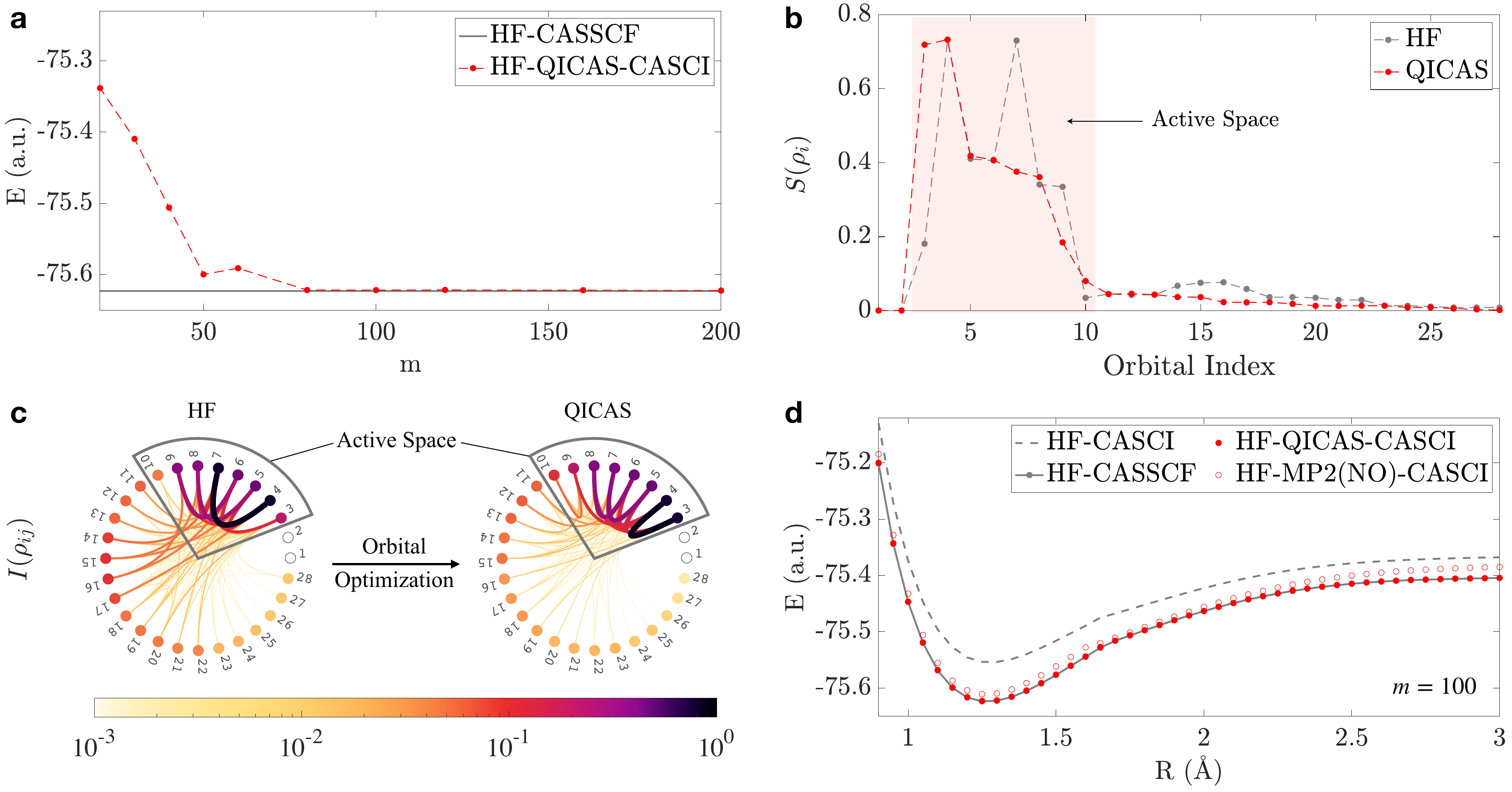}
    \caption{\textbf{a}. HF-QICAS-CASCI(8,\:8) energy for $\mathrm{C_2}$ in the cc-pVDZ basis at $\mathrm{R}=1.243\,\mathrm{\AA}$ versus the bond dimension $m$ of the MPS ground state $|\Psi_0\rangle$. The HF-CASSCF energy is plotted as a reference. \textbf{b}. Single orbital entropy $S(\rho_i)$ and \textbf{c}. orbital-orbital correlation $I(\rho_{ij})$ of the Hartree-Fock orbitals (HF) and QICAS optimized orbitals (QICAS) for $\mathrm{C_2}$ in the cc-pVDZ basis set, with internuclear distance at $\mathrm{R}=1.243\,\mathrm{\AA}$. \textbf{d}. HF-CASCI(8,\:8), HF-CASSCF(8,\:8), HF-QICAS-CASCI(8,\:8), and HF-MP2(NO)-CASCI(8,\:8) energy (a.u.) for $\mathrm{C}_2$ with cc-pVDZ basis set as functions of internuclear distance R ($\mathrm{\AA}$). See text for more details.}
    \label{fig:qicas_c2_combined}
\end{figure*}

We first discuss application (\textbf{A}) which is particularly suitable for smaller systems of moderate complexity such as, e.g., $\mathrm{C_2}$ in the cc-pVDZ basis set at equilibrium ($\mathrm{R}=1.243\,\mathrm{\AA}$\cite{huber2013molecular}).
One important observation we made during the implementation is that the efficacy of QICAS depends, up to a point, on the bond dimension $m$ of the MPS ground state $|\Psi_0\rangle$. With a higher bond dimension $m$, the multireference character of the true ground state is captured more accurately for the subsequent QICAS analysis, but a higher computational cost entails. In Figure \ref{fig:qicas_c2_combined} \textbf{a} we present the HF-QICAS-CASCI energy $E_0^\mathrm{CASCI}(\mathcal{B}^\ast)$ in a CAS(8,\:8) against the bond dimension $m$ while keeping the number of DMRG sweeps fixed at 50. In accordance with our intuition, a higher $m$ results in a better set of optimized orbitals and a lower CASCI energy. With an error of 1.4 mHa for $m=80$, our analysis confirms that QICAS identifies highly accurate active spaces
for bond dimensions $m\geq80$ and the respective post-QICAS CASCI energies are within chemical accuracy ($<$ 1.6 mHa) of the HF-CASSCF energies. 

To understand better the mechanism of QICAS, we compare the orbital entanglement structure in the HF ($\mathcal{B}_{\mathrm{HF}}$) and QICAS optimized basis ($\mathcal{B}^\ast$). For this, we analyze the $m=200$ data point from Figure \ref{fig:qicas_c2_combined} (a), where the MPS ground state captured the most orbital correlation.
In Figure \ref{fig:qicas_c2_combined} \textbf{b}, we present the single orbital entropies $S(\rho_i)$ for both bases. We recall that the single orbital entropies of the inactive orbitals explicitly enter the cost function $F_\mathrm{QI}(\mathcal{B})$. For the non-active orbitals, the entropy profile in the QICAS optimized basis is much lower as we expected. Most of the reduction occurs for orbitals 14-17, which are considerably correlated in the HF basis. The out-of-CAS orbital correlation $F_\mathrm{QI}(\mathcal{B})$ reduces from 0.64 in the HF basis $\mathcal{B}_{\mathrm{HF}}$ to 0.38 in the QICAS optimized basis $\mathcal{B}^\ast$. For the active orbitals, although the two profiles look rather similar up to some reordering, the sum of orbital entropies is actually slightly higher in the QICAS basis, namely 3.28 compared to 3.17 in the HF basis. This might seem counter-intuitive at first, as we aim to simplify and not further complicate the orbital correlation of the system, but it in fact aligns perfectly with the purpose of QICAS: It determines an active space that discards the \emph{least} amount of correlation by reducing the orbital correlation within the inactive space, while simultaneously shuffling some excess orbital correlation into the active orbitals.

With this observation in mind,
we remark that the QICAS optimized single orbital entropy profile
may be able to assist and upgrade the procedure of active space \textit{size} selection in Ref.~\cite{reiher16}.
For an initial proposal, one could minimize the sum of \textit{all} orbital entropy (an inexpensive subroutine after obtaining the 1- and 2-RDM),
and use the unbiased optimized basis for predicting an active space size $D_{\mathrm{CAS}}$ based on the plateau structure of the orbital entropy.
$D_{\mathrm{CAS}}$ is then fed into the QICAS routine.
As active space size selection is not the main focus of this work,
interested readers can refer to Appendix \ref{app:size} for more details.


Besides the improvement in the single orbital entropy profile, the success of QICAS lies also in the consistent simplification of the correlation pattern \textit{between} orbitals. To verify this on a quantitative level, we measure the correlation between any two orbitals $i$ and $j$ with the quantum mutual information\cite{MI1,legeza03,MI2,rissler06orbital,MI3,reiher12entanglement,boguslawski2015orbital,ding2020concept}
\begin{equation}
    I(\rho_{ij})\equiv S(\rho_i)+S(\rho_j) - S(\rho_{ij}). \label{eqn:orb-orb}
\end{equation}
It is worth noting that the orbital-orbital mutual information \eqref{eqn:orb-orb}
has also been used for active space size selection in a fixed orbital basis\cite{boguslawski2017multi},
in a similar fashion to Ref.~\cite{reiher16}.
However, for the purpose of orbital \textit{optimization} as in QICAS, \eqref{eqn:orb-orb} is far too costly, as it involves up to the 4-RDM.
In our calculation, the two-orbital reduced density matrices $\rho_{ij}$ are obtained by using the \textsc{Syten} toolkit\cite{hubig:_syten_toolk,hubig17:_symmet_protec_tensor_networ}.
In Figure \ref{fig:qicas_c2_combined} \textbf{c} we present $I(\rho_{ij})$ for both bases as colored curves connecting corresponding vertices $i,j$, with the single orbital entropies $S(\rho_i)$ color-coded on the respective vertices. Compared to the HF basis, the ``correlation net'' is indeed thinner for the QICAS optimized basis. In particular, the pairwise correlation associated with orbitals 14-17 is absorbed into the active space. Figure \ref{fig:qicas_c2_combined} \textbf{b} and \textbf{c} together unambiguously demonstrate the ability of QICAS in positioning most of the correlation effects into the active space, thus achieving a more compact active space structure, and in virtue of Theorem \ref{thrm} an optimal CASCI energy.

\begin{figure*}[htb]
    \centering
    \includegraphics[scale=0.3]{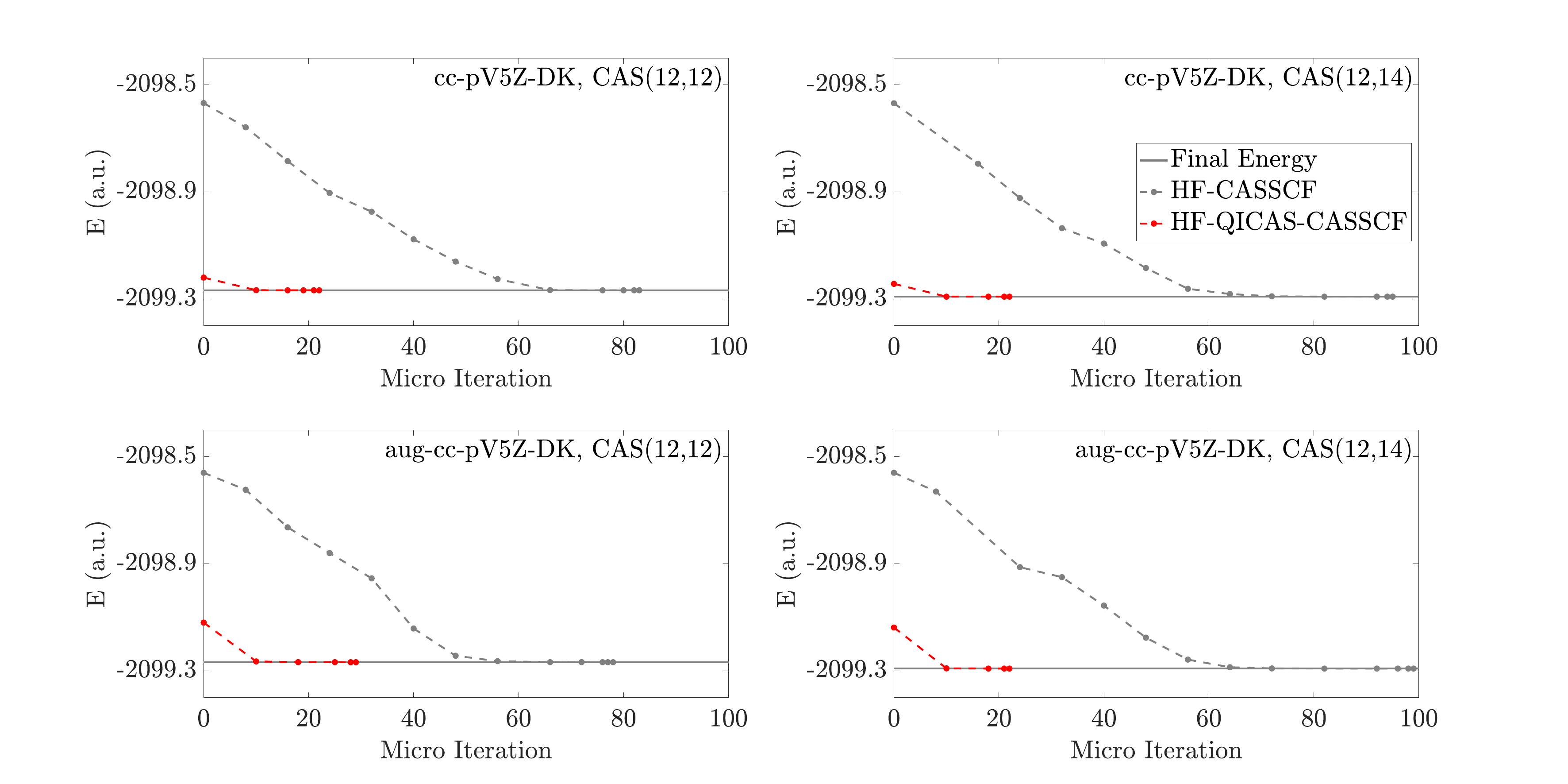}
    \caption{Energy evolution as the number of micro iterations in CASSCF(12,12) and CASSCF(12,14) for $\mathrm{Cr_2}$ in the cc-pV5Z-DK and aug-cc-pV5Z-DK basis set at $\mathrm{R}=1.679\,\mathrm{\AA}$, starting from HF-orbitals and from QICAS optimized orbitals. See text for more details.} 
    \label{fig:qicas_cr2_combined}
\end{figure*}

The nearly perfect efficacy of QICAS indeed holds for all internuclear distances of $\mathrm{C_2}$. In Figure \ref{fig:qicas_c2_combined} \textbf{d} we present one of our key results: using the HF-QICAS-CASCI method we are able to reproduce the entire dissociation curve of $\mathrm{C_2}$ in the cc-pVDZ basis with an error compared to the HF-CASSCF method of up to only 2.3 mHa. For most of the dissociation curve (for 39 out of 41 evenly spaced data points), the error is within chemical accuracy ($<$ 1.6 mHa), as evidenced by the fact that the HF-QICAS-CASCI data points (red dots) are virtually on top of the HF-CASSCF ones (grey dots on solid line). Tabulated data can be found in Appendix \ref{app:data} obtained by performing QICAS based on MPS ground states with maximal bond dimension $m=100$ and 50 QC-DMRG sweeps. The HF-CASSCF energy at $\mathrm{R}=1.65\mathrm{\AA}$ is computed with the \textsc{Molpro}\cite{molpro1,molpro2,molpro3} package, and all other HF-CASCI/CASSCF energies with \textsc{PySCF}\cite{sun2018pyscf}. Here we again emphasize that the orbital optimization is based solely on the quantum information-inspired cost function $F_\mathrm{QI}(\mathcal{B})$ in \eqref{eqn:cost_QICAS}, which, although is not directly connected to the highly involved CASCI energy landscape, leads to the correct minimum. The success of QICAS confirms that the underlying correlation structure of the approximate ground state $|\Psi_0\rangle$ can indeed be exploited for an accurate active space identification. Moreover, the useful part of the correlation is mostly static, evident by the fact that even when $|\Psi_0\rangle$ is approximated by an MPS with small bond dimension $m=100$, the correlation captured by the MPS is already sufficient for QICAS to identify a chemically accurate active space.
Lastly, we compared in Figure \ref{fig:qicas_c2_combined} \textbf{d} the CASCI energy between the QICAS optimized basis and the natural orbital (NO) basis of the second-order M{\o}ller-Plesset perturbation theory (MP2) solution.
In this example we find the QICAS orbitals to be consistently better.
At the same time, QICAS does not suffer from the potential failure of MP2 when degeneracies arise due to dissociation.

For larger systems, performing a QC-DMRG calculation involving \emph{all} orbitals can be rather expensive, if not unrealistic, even for moderate bond dimensions. To overcome this limitation, we propose to optimize just a subset $\mathcal{S}$ of $\Tilde{D}=|\mathcal{S}|$ orbitals with QICAS, based on the MPS ground state on the chosen $\Tilde{D}$ orbitals.
Since not all orbitals in the basis are optimized, a gap would inevitably appear between the HF-QICAS-CASCI energy and the HF-CASSCF energy. Nonetheless, one would expect the former to be still much lower than the HF-CASCI energy, and that the convergence of CASSCF could be accelerated by starting with the optimized basis $\mathcal{B}^\ast$ instead of the HF orbitals.

To showcase exactly this second promising application (\textbf{B}) of QICAS, we study the notoriously strongly correlated $\mathrm{Cr_2}$ at an internuclear distance $\mathrm{R}=1.679\,\mathrm{\AA}$ around the experimental equilibrium\cite{bondybey1983electronic}. The cc-pV5Z-DK (306 orbitals) and aug-cc-pV5Z-DK (404 orbitals) basis sets\cite{balabanov2005systematically} are used, together with the scalar relativistic exact two-component (X2C) Hamiltonian\cite{kutzelnigg2005quasirelativistic,peng2012exact}. With such large basis set sizes, CASSCF starts to become costly, even with a small active space for which full CI is still feasible. We consider in the following two active space sizes, CAS(12,\:12) and CAS(12,\:14) and freeze the most inner 12 HF orbitals (the ``$\mathrm{Mg}$ core''). We reduce the scope of QICAS to the subset of $\Tilde{D}=60$ lowest lying HF orbitals above the fixed $\mathrm{Mg}$ core, and correlate 24 electrons, such that sufficient electron correlation can be captured at a manageable cost. The bond dimension of the 60-orbital MPS is set at $m=250$ and 50 DMRG sweeps are performed. 

\begin{table*}
    \centering
    \begin{tabular}{llllllll}
    \hline \rule{0pt}{3ex}\rule[-1.5ex]{0pt}{0pt}Basis & $D$ & Method($N_\mathrm{CAS}$,\:$
    D_\mathrm{CAS}$) & Energy (a.u.) $\quad\quad$ & Post-HF Time (h) $\quad$ & $\#$ Micro $\quad$ & $\#$ JK
     \\
     \hline
     \hline
    cc-pV5Z-DK & 306 $\quad$ & HF-CASSCF(12,\:12) $\quad$ & -2099.26725956 & 1.24  & 83 & 358 \rule{0pt}{4ex}\rule[-1.5ex]{0pt}{0pt}
    \\
              & & HF-QICAS-CASSCF(12,\:12) $\quad$ & -2099.26725956 & 1.57 (1.25+0.32)  & 22 & 60 \rule{0pt}{0ex}\rule[-1.5ex]{0pt}{0pt}
    \\
     &  $\quad$ & HF-CASSCF(12,\:14) $\quad$ & -2099.29109845 & 8.24  & 95 & 408 \rule{0pt}{4ex}\rule[-1.5ex]{0pt}{0pt}
    \\
              & & HF-QICAS-CASSCF(12,\:14) $\quad$ & -2099.29109856 & 3.18 (1.40+1.78)  & 22 & 55 \rule{0pt}{0ex}\rule[-1.5ex]{0pt}{0pt}
    \\
    aug-cc-pV5Z-DK $\quad$ & 404 $\quad$ & HF-CASSCF(12,\:12) $\quad$ & -2099.26745743 & 2.93 & 78 & 328 \rule{0pt}{4ex}\rule[-1.5ex]{0pt}{0pt}
    \\
      &  & HF-QICAS-CASSCF(12,\:12) $\quad$ & -2099.26745742 & 1.99 (1.04+0.95) & 29 & 88 \rule{0pt}{0ex}\rule[-1.5ex]{0pt}{0pt}
    \\
              & & HF-CASSCF(12,\:14) $\quad$ & -2099.29130834 & 9.46 & 99 & 415 \rule{0pt}{4ex}\rule[-1.5ex]{0pt}{0pt}
    \\
              & & HF-QICAS-CASSCF(12,\:14) $\quad$ & -2099.29130846 & 3.38 (1.13+2.25) & 22 & 65 \rule{0pt}{0ex}\rule[-2ex]{0pt}{0pt}
    \\
    \hline
    \end{tabular}
    \caption{Computational cost comparison of CASSCF(12,12) and CASSCF(12,14) for $\mathrm{Cr_2}$ in the cc-pV5Z-DK and aug-cc-pV5Z-DK basis set at $\mathrm{R}=1.679\,\mathrm{\AA}$, between starting from HF-orbitals and from QICAS optimized orbitals. In particular, the post-HF time when QICAS is invoked is broken down into the run time of QICAS and that of the CASSCF calculation.} 
    \label{tab:performance}
\end{table*}

To compare the convergence of CASSCF starting from either the HF or QICAS optimized orbitals, we present in Figure \ref{fig:qicas_cr2_combined} the respective energy evolution against the number of CASSCF micro iterations.
The horizontal gap between two adjacent data points represents one macro iteration\footnote{In the case of CAS(12,14), for both basis sets the state drifted to a subspace of incorrect spin symmetry at one macro iteration during the HF-CASSCF calculation. Both macro iterations are omitted from the plot, but the exact values of the unphysical drifted energy can be found in Appendix \ref{app:data}. The spin symmetry remains unbroken in all other macro iterations, and is retained in all final CASSCF solutions.}. Other indicators of computational effort of CASSCF are displayed in Table \ref{tab:performance}, including the number of micro iteration (\# micro), the number of times the two-electron integrals \textbf{J} and \textbf{K} are computed (\# JK), and the post-HF run time (all calculations are performed on an Intel Xeon Platinum 8358 central processing unit).
The results clearly reveal that QICAS provides a distinctly superior starting basis for CASSCF and can significantly accelerate convergence. For both basis sets, and for both active space sizes, the HF-QICAS-CASCI energy (at 0 iteration in the respective plots) is already much lower than the HF-CASCI energy, demonstrating a drastic improvement thanks to QICAS relative to the HF orbitals. Furthermore, starting from the optimized orbitals the correct CASSCF energy is practically reached after only 1 or 2 macro iterations. In contrast, starting from HF-orbitals CASSCF takes at least 8 macro iterations to reach chemical accuracy. 
Consequently, each post-QICAS CASSCF calculation requires much fewer computations of the two-electron integrals, and the run time is overall 3-4 times shorter than that of the corresponding post-HF CASSCF calculation.

Obtaining the QICAS optimized orbitals is certainly not for free. Yet, the effort invested in computing the MPS ground state and minimizing the out-of-CAS orbital correlation $F_\mathrm{QI}(\mathcal{B})$ is well exceeded by the computational cost saved in a post-QICAS CASSCF.
The bulk of the QICAS run time is spent on the QC-DMRG calculation, whereas the minimization of $F_\mathrm{QI}(\mathcal{B})$ takes only a few minutes. In this sense, QICAS is at least as economic as other methods that are based as well on MPS ground state approximations with low bond dimension (as, e.g., Ref.~\cite{reiher16}), while at the same time QICAS offers an advantageous starting point for CASSCF uniquely achieved by our QI-motivated orbital optimization.

To conclude, in the form of QICAS we proposed an optimization scheme which exploits through concise quantum information (QI) tools the orbital correlation structure of cheaper post-HF solutions to determine accurate active spaces.
For this, based on a preceding QC-DMRG calculation with low bond dimension, QICAS optimizes the orbital basis $\mathcal{B}$ by minimizing the sum of orbital correlations $F_\mathrm{QI}(\mathcal{B})$ of the non-active orbitals. According to Theorem \ref{thrm}, this choice of a cost function is very well-motivated:  $F_\mathrm{QI}(\mathcal{B})$ bounds exactly that portion of the correlation energy which the corresponding active space fails to recover. In practical applications QICAS demonstrated a remarkable potential. For $\mathrm{C_2}$, we observed an excellent agreement between the CASCI energy of the QICAS optimized orbitals and the CASSCF energy along the entire dissociation curve. For the more challenging $\mathrm{Cr_2}$, we found that optimizing a subset of orbitals using QICAS prepares a superior starting point for CASSCF, and consequently greatly reduces the number of iterations required for convergence.

From a more general and conceptual point of view, our work establishes for the first time a pivotal rigorous connection between orbital correlation structures and the quality of corresponding active spaces quantified in terms of CASCI energies. Supported by this key result, our findings suggest a change of paradigm in addressing the ground state problem: QICAS could be added as an intermediate layer between initial HF computation and final post HF-treatment to improve the overall accuracy and efficiency.

Last but not least, we believe that the potential transformative impact of QI tools in theoretical and computational chemistry to simplify electronic structure is not limited to active space methods. Schemes and ideas related to QICAS may lead to advances in the development and application of any method where orbital optimization plays a prominent role. Prime examples would be quantum computing\cite{vqe1,vqe2,vqe3,qcqc1}, where a poor choice of reference orbitals results in a large circuit depth, and the density matrix embedding theory\cite{dmet1,dmet2,dmet3}, where the sought-after local fragment and bath orbitals form a structure similar to that of an active space\cite{dmet2}.

\begin{acknowledgements}
We thank Peter Knowles for helpful discussions and acknowledge financial support from the Deutsche Forschungsgemeinschaft (DFG, German Research Foundation), Grant SCHI 1476/1-1 (L.D., C.S.), and the Munich Center for Quantum Science and Technology (L.D., C.S.). The project/research is also part of the Munich Quantum Valley, which is supported by the Bavarian state government with funds from the Hightech Agenda Bayern Plus.
\end{acknowledgements}

\bibliography{refs}

\appendix

\section{Proof to Theorem 1} \label{app:proof}

In this section we provide a proof to Theorem \ref{thrm} in the Letter.

\begin{proof}
For a fixed orbital reference basis $\mathcal{B}$, and a CAS($N_\mathrm{CAS}$,\:$D_\mathrm{CAS}$), the corresponding linear space of complete active space $N$-electron wavefunctions is given by
\begin{equation}
\begin{split}
    \mathcal{S}_\text{CAS} \equiv \{|\Psi\rangle =|\mathbf{n}\rangle_\mathcal{N}&\otimes|\psi\rangle_\mathcal{A}, \text{ with } \hat{N}|\Psi\rangle = N|\Psi\rangle,
    \\
    \hat{N}|\psi\rangle_A &= N_\mathrm{CAS}|\Psi\rangle\}.
\end{split}
\end{equation}
Here $|\mathbf{n}\rangle_\mathcal{N} = \bigotimes_{i\in\mathcal{N}}|n_i\rangle$, where $n_i \in \{0,\uparrow\downarrow\}$ defines the occupancies of the non-active orbitals $\phi_i$, $i \in \mathcal{N}$, $\hat{N}$ denotes the total particle number operator, and $|\psi\rangle_\mathcal{A}$ is an $N_\mathrm{CAS}$-electron wavefunction defined within the active space of $D_\mathrm{CAS}$ active orbitals. The orthogonal projection operator onto the subspace $\mathcal{S}_\text{CAS}$ shall be denoted by $\hat{P}$. In the following, we consider two states in $\mathcal{S}_\text{CAS}$, the state $|\Psi_\text{CAS}'\rangle$ which has maximal overlap with the full CI ground state $|\Psi_0\rangle$,
\begin{equation}
    |\Psi_\text{CAS}'\rangle = \arg\max_{|\Phi\rangle \in \mathcal{S}_\text{CAS}}|\langle \Phi|\Psi_0\rangle|^2 = \frac{\hat{P}|\Psi_0\rangle}{\|\hat{P}|\Psi_0\rangle\|_2},
\end{equation}
and the $\mathcal{B}$-CASCI ground state $|\Psi_\text{CAS}\rangle$ of the Hamiltonian $\hat{H}$,
\begin{equation}
    |\Psi_\text{CAS}\rangle = \arg\min_{|\Phi\rangle \in \mathcal{S}_\text{CAS}} \langle \Phi | \hat{H} |\Phi \rangle.
\end{equation}
Then, in virtue of the Rayleigh-Ritz variational principle, we have $\langle \Psi_\text{CAS}' |\hat{H}|\Psi_\text{CAS}'\rangle \geq \langle \Psi_\text{CAS}|\hat{H}|\Psi_\text{CAS}\rangle$. Using the spectral decomposition of the Hamiltonian $\hat{H} = \sum_{i\geq 0} E_i |\Psi_i\rangle \langle \Psi_i|$, we can split the former as
\begin{equation}
\begin{split}
    E' &\equiv \langle \Psi_\text{CAS}'|\hat{H}|\Psi_\text{CAS}'\rangle
    \\
    & = E_0 |\langle\Psi_\text{CAS}'| \Psi_0\rangle|^2 + \sum_{i > 0} E_i |\langle\Psi_\text{CAS}'| \Psi_i \rangle|^2
    \\
    & \leq E_0 |\langle \Psi_\text{CAS}'| \Psi_0 \rangle|^2 + \max_i E_i (1-|\langle \Psi_\text{CAS}'|\Psi_0 \rangle|^2).
\end{split}
\end{equation}
By introducing $\epsilon \equiv 1-|\langle \Psi_\text{CAS}'|\Psi_0\rangle|^2$, we then obtain
\begin{equation}
    \Delta E \leq \Delta E' \leq \Delta E_\text{max} \epsilon, \label{eqn:a5}
\end{equation}
where $\Delta E' \equiv E' - E_0$, and $\Delta E_{\max} \equiv \max_i{E_i}-E_0$.

Since $|\Psi_\mathrm{CAS}'\rangle = \hat{P} |\Psi_0\rangle/\sqrt{1-\epsilon}$, we can rewrite the full CI ground state as
\begin{equation}
\begin{split}
    |\Psi_0\rangle &= \hat{P}|\Psi_0\rangle + (\openone - \hat{P}) |\Psi_0\rangle
    \\
    &= \sqrt{1-\epsilon} |\Psi_\mathrm{CAS}'\rangle + \sqrt{\epsilon} \sum_{\mathbf{m}\neq \mathbf{n}} \lambda_\mathbf{m} |\mathbf{m}\rangle \otimes |\psi_\mathbf{\mathbf{m}}\rangle,
\end{split}
\end{equation}
where $\sum_{\textbf{m}\neq\textbf{n}}|\lambda_\textbf{m}|^2 = 1$.
Using the fact that $\rho_i$ is diagonal in the occupation number basis (as a consequence of fixed particle number and magnetization of the system), we can easily derive the orbital reduced density matrices of the non-active orbitals. For example, if orbital $i$ is virtual, then the diagonal elements of $\rho_i$ are given by
\begin{equation}
    \begin{split}
        (\rho_i)_{00} &= 1-\epsilon + \epsilon \sum_{\textbf{m}\neq\textbf{n}:m_i = 0} |\lambda_\textbf{m}|^2 \equiv 1-\epsilon p_i,
        \\
        (\rho_i)_{11} &= \epsilon \sum_{\textbf{m}\neq\textbf{n}:m_i = \uparrow} |\lambda_\textbf{m}|^2 \equiv \epsilon q_i,
        \\
        (\rho_i)_{22} &= \epsilon \sum_{\textbf{m}\neq\textbf{n}:m_i = \downarrow} |\lambda_\textbf{m}|^2 \equiv \epsilon r_i,
        \\
        (\rho_i)_{33} &= \epsilon \sum_{\textbf{m}\neq\textbf{n}:m_i = \uparrow\downarrow} |\lambda_\textbf{m}|^2 \equiv \epsilon s_i,
    \end{split}
\end{equation}
where $p_i = \sum_{\textbf{m} \neq \textbf{n}:m_i \neq 0} |\lambda_{\textbf{m}}|^2$. Analogous expressions can be derived for frozen core orbitals. It follows that $p_i = q_i + r_i + s_i$ and thus $\bm{y} \equiv (1-\epsilon p_i,\epsilon p_i,0,0) \succ \bm{x} \equiv (1- \epsilon p_i, \epsilon q_i, \epsilon r_i, \epsilon s_i)$, where $\succ$ stands for vector majorization. We can then bound the sum of entropies $S(\rho_i)$ of the non-active orbitals, using the Schur concavity of the Shannon entropy $H(\bm{x}) = -\sum_i x_i \log(x_i)$, as
\begin{equation}
    \begin{split}
        S(\rho_i) = H(\bm{x}) \geq H(\bm{y}) = B(\epsilon p_i).
    \end{split}
\end{equation}
Here, we $B(x) \equiv -x\log(x)-(1-x)\log(1-x)$ denotes  the binary entropy.
By assuming $\epsilon < 1/2$, which is easily fulfilled as long as $\mathcal{B}$-CASCI approximates the true ground state reasonably well, we further obtain
\begin{equation}
    \sum_{i\in \mathcal{N}} S(\rho_i) \geq \log(4)\epsilon \sum_{i\in\mathcal{N}} p_i \geq \log(4) \epsilon.
\end{equation}
Here, we used the convexity of the binary entropy and the fact that every $|\lambda_{\textbf{m}}|^2$ occurs at least once in the sum $\sum_{i\in\mathcal{N}} p_i$. Finally, using \eqref{eqn:a5} we arrive at the result presented in the theorem, i.e.,
\begin{equation}
    \Delta E \leq \frac{\Delta E_\text{max}}{\log(4)} \sum_{i\in\mathcal{N}} S(\rho_i) = \frac{\Delta E_\text{max}}{\log(4)}  F_\mathrm{QI}(\mathcal{B}).
\end{equation}
In other words, when the sum $F_\mathrm{QI}(\mathcal{B})$ of entanglement entropies of the non-active orbitals is sufficiently small, the $\mathcal{B}$-CASCI energy approximates well the full CI ground state energy.
\end{proof}

\section{Orbital Optimization} \label{app:opt}

An essential step in the QICAS subroutine is to minimize the cost function $F_\mathrm{QI}(\mathcal{B})$, which only involves the input 1- and 2-RDM, $\bm{\gamma}$ and $\bm{\Gamma}$. In the latter, only the entries $\Gamma^{i\bar{j}}_{k\bar{l}}$ are needed, where $i,j,k,l$ are spatial orbital indices whereas the absence (presence) of a bar indicates the $\uparrow$ ($\downarrow$) spin. These relevant entries of the 2-RDM can be collected as a smaller tensor $\bm\Gamma_\mathrm{rel}$, which contains only 1/16 of the entries of the full 2-RDM $\bm\Gamma$.

The goal is now to find the orbital basis $\mathcal{B}^\ast$ for which $F_\mathrm{QI}(\mathcal{B})$ is minimal. An orthogonal transformation $\mathbf{U}=(U_{ij})_{ij}$ transforms the basis $\mathcal{B} = \{\phi_i\}$ to $\Tilde{\mathcal{B}} = \{\tilde\phi_i\}$ according to
\begin{equation}
    \tilde\phi_i = \sum_j U_{ij} \phi_j.
\end{equation}
Using the HF-orbitals $\mathcal{B}_\mathrm{HF} = \{\phi^{(\mathrm{HF})}_i\}$ as a standard starting point, we can identify each ordered basis $\mathcal{B}$ with the transformation $\mathbf{U}$ from $\mathcal{B}_\mathrm{HF}$ to $\mathcal{B}$ and thus
\begin{equation}
    F_\mathrm{QI}(\mathcal{B}) \equiv F_\mathrm{QI}(\mathbf{U}).
\end{equation}
The orthogonal matrix can be parameterized as
\begin{equation}
\mathbf{U} \equiv \exp(\mathbf{X}) = \openone + \mathbf{X} + \frac{1}{2} \mathbf{X}^2 + \cdots,
\end{equation}
with an anti-symmetric matrix $\mathbf{X}$ of the same size $D\times D$ satisfying $\mathbf{X}^\mathrm{T} = -\mathbf{X}$. Now the problem of finding the minimizing basis $\mathcal{B}^\ast$ is transformed to optimizing the $D(D-1)/2$ independent parameters $X_{ij}$'s in $\mathbf{X}$.

Here we would like to point out that, although $F_\mathrm{QI}(\mathcal{B})$ only involves the non-active indices, it is not invariant under internal rotation within the non-active space. This is a major difference from the orbital optimization in CASSCF, where the energy is invariant under rotation within the active, closed, and virtual orbital spaces, separately. In theory, to find the true minimum of $F_\mathrm{QI}(\mathcal{B})$, one should optimize the parameters $X_{ij}$ connecting the active and non-active indices, as well as the parameters with both indices belonging to either closed or virtual orbitals. For the $\mathrm{C_2}$ molecule, as we are aiming for chemical accuracy with post-QICAS CASCI, we taken into account all relevant indices. But for the scenario where QICAS is used to prepare an advantageous starting point for CASSCF, we found that optimizing only the parameters $X_{ij}$ connecting the active and non-active orbitals does not qualitatively hamper the performance. Moreover, omitting the rotations between non-active orbitals saves considerable amount of computational time, especially when the underlying basis set is much larger than the size of the active space. This more economic approach is adopted for the $\mathrm{Cr_2}$ molecule in the main text.

For the optimization we employed a gradient-free method: each parameter $X_{ij}$ is optimized individually, and after optimizing each $X_{ij}$ the 1- and 2-RDM are updated and the history of each two-orbital rotation is documented. The final orthogonal matrix $\mathbf{U}^\ast$ is realized by a sequence of Jacobi rotations $\mathbf{T}_{ij}(X_{ij})$ between the orbital $i$ and $j$ characterized by the optimized rotation angle $X_{ij}$
\begin{equation}
    \mathbf{U}^\ast = \prod_{n=1}^{N_\mathrm{cycle}} \prod_{(i,j)} T_{ij}(X_{ij}^{(n)}) \mathbf{U}_0.
\end{equation}
Here, $N_\mathrm{cycle}$ is the number of times the list of relevant parameters is repeated, and $\mathbf{U}_0$ is the starting unitary which can either be the identity or randomized. The process of obtaining $\mathbf{U}^\ast$ can be summarized as follows:
\begin{enumerate}
    \item Shuffle the orbital indices $L = \{1,2,\ldots,D\}$ and denote it as $L' = \{d_1,d_2,\ldots, d_D\}$.

    \item For $i,j \in L'$, if $i \in \mathcal{N}$ or $j \in \mathcal{N}$ (to ignore rotations within the non-active space, replace or with the exclusive or), do the next step.

    \item For $X_{ij}\in[0,\pi]$ in step size $10^{-2}$, evaluate $F_\mathrm{QI}(\mathbf{T}_{ij}(X_{ij}))$ and find the minimizing $X_{ij}^\ast$. Then in the smaller interval $[X_{ij}^\ast-10^{-2}, X_{ij}^\ast+10^{-2}]$, in step size $10^{-4}$ sample again the minimum of $F_\mathrm{QI}(\mathbf{T}_{ij}(X_{ij}))$. If the improvement $F_\mathrm{QI}(\mathbf{0})-F_\mathrm{QI}(\mathbf{T}_{ij}(X_{ij})) > \epsilon_1$, rotate $\bm\gamma$ and $\bm\Gamma$ by $\mathbf{T}_{ij}(X_{ij})$, and update $\mathbf{U}$ with $\mathbf{T}_{ij}(X_{ij}) \mathbf{U}$.

    \item For $N_\mathrm{cycle}$ times, repeat the above procedures, or until the improvement in $F_\mathrm{QI}$ after the cycle is less than $\epsilon_2$ compared to the end of the last cycle. When the algorithm terminates, the optimized orthogonal matrix is then $\mathbf{U}^\ast = \mathbf{U}$.
\end{enumerate}
For our calculations, we set $\epsilon_2 = 10\epsilon_1$. In the $\mathrm{C_2}$ example, $\epsilon_1$ is set to $10^{-8}$. In the $\mathrm{Cr_2}$ example, $\epsilon_1$ is set to $10^{-7}$. The number of cycles is set at $N_\mathrm{cycle} = 200$. Due to the stochastic nature of the partially converged DMRG calculation and the optimization of the cost function, for the HF-QICAS-CASCI data in Figure 3 \textbf{d} the calculation is repeated 50 times and the lowest CASCI energy is taken. For $\mathrm{Cr_2}$, as we opted not to optimize within the non-active subspace, convergence of the optimization of $F_\mathrm{QI}$ is reached with not more than 10 cycles in all calculations.

\section{Determining Active Space Size with QICAS} \label{app:size}

\begin{figure}
    \centering
    \includegraphics[scale=0.3]{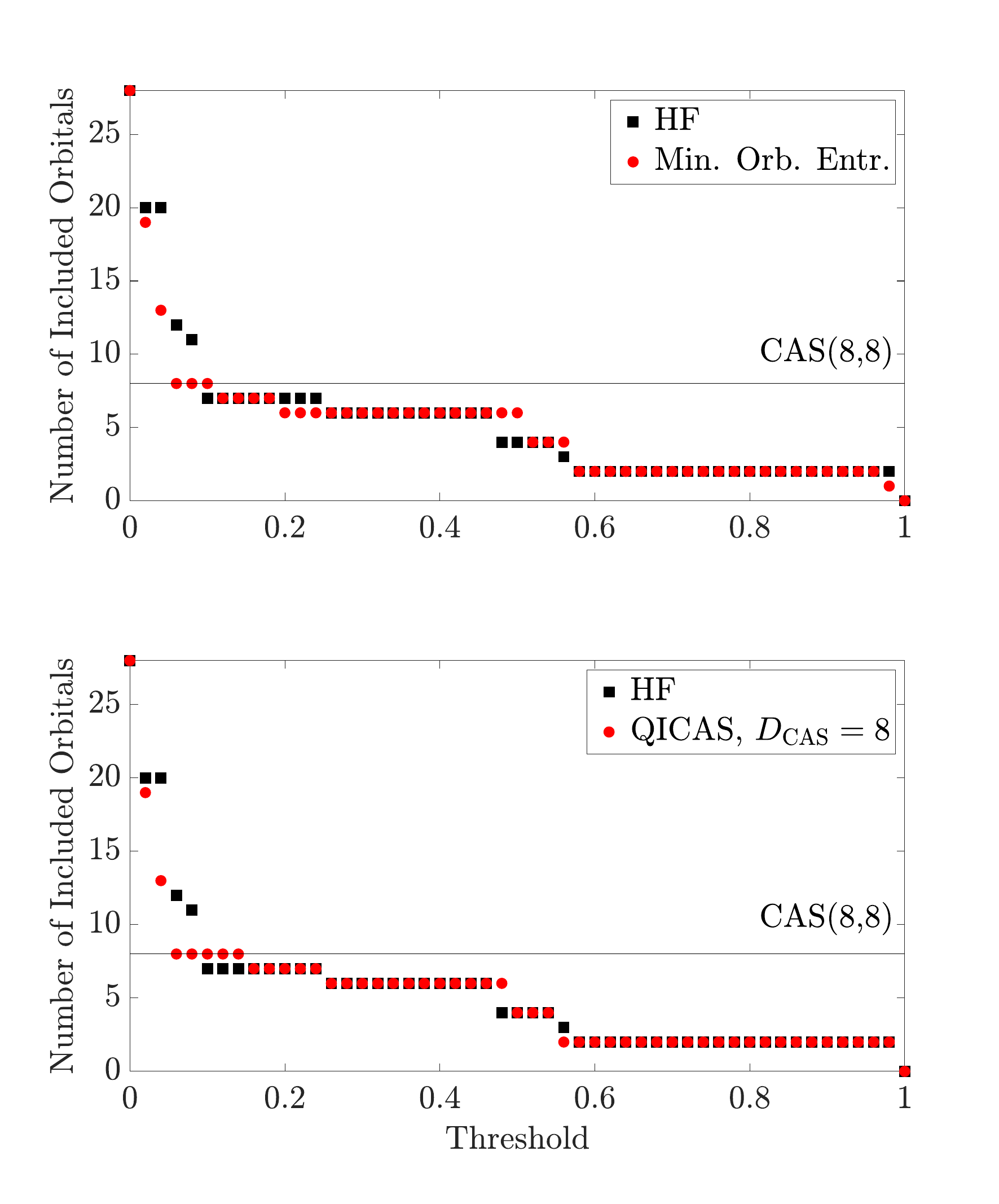}
    \caption{Threshold diagrams of the entropy of three orbital basis for the $\mathrm{C_2}$ molecule at equilibrium geometry $R=1.243\mathrm{\AA}$ in cc-pVDZ basis set:
    the Hartree-Fock (HF) basis as the initial guess, the basis with minimal overall orbital entropy (Min. Orb. Entr.) which minimizes $F'_\mathrm{QI}(\mathcal{B})$,
    and the QICAS optimized basis with $D_{\mathrm{CAS}} = 8$. }
    \label{fig:threshold}
\end{figure}

An apparent requirement for implementing the orbital optimization for the minimization of the QI cost function $F_\mathrm{QI}(\mathcal{B})$ is the number of active/non-active orbitals.
This feature is however not a drawback, as we readily have all the ingredients to calculate the entropy profile of the initial set of orbitals once the 1- and 2-RDMs are obtained,
which can be used for active space \textit{size} selection, also known as AutoCAS proposed by Stein and Reiher\cite{reiher16}.
The main idea in Ref.~\cite{reiher16} is that the set of orbitals with substantially higher orbital entropy than the others should constitute the active space.
To be more concrete, Stein and Reiher utilized the so-called threshold diagrams,
where they count for the number of orbitals whose entropy exceed a series percentage of the maximal entropy among all orbitals.
This diagram effects a top-down scanning of the entropy profile for identifying strong orbital correlation.

The success of AutoCAS relies on two assumptions:
(i) the optimal active orbitals exhibit a high plateau in the orbital entropy profile,
and (ii) the entropy profile of the current basis set based on which the prediction of the active space size is made, is similar to that of the optimal basis.
Assumption (i) lies in the heart of almost all correlation-based active space selection methods.
Instead of high orbital entropy, active orbitals can also distinct themselves with other indicators of strong orbital correlation such as fractional occupation numbers.
The second assumption, however, may not always hold.
A poorly chosen set of initial guess orbitals could have a qualitatively different entropy profile than that of the optimal orbitals.

QICAS can lift the assumption (ii), by optimizing the orbital entropy profile.
For this, we propose to generalize the QICAS cost function $F_{\mathrm{QI}}(\mathcal{B})$ to the following
\begin{equation}
    F'_{\mathrm{QI}}(\mathcal{B}) = \sum_i S(\rho_i),
\end{equation}
namely summing over all orbitals instead of just the non-active ones.
This yields an unbiased optimization of the overall orbital entropy, based on which a threshold diagram can be produced.
Once an active space size $D_\mathrm{CAS}$ is predicted, an ordinary QICAS routine can be run.

As an example, we present in Figure \ref{fig:threshold} the threshold diagrams of the entropy of three orbital bases for the $\mathrm{C_2}$ molecule at equilibrium geometry $R=1.243\mathrm{\AA}$ in cc-pVDZ basis set:
(a) the Hartree-Fock (HF) basis as the initial guess, (b) the basis with minimal overall orbital entropy (Min.~Orb.~Entr.~) which minimizes $F'_\mathrm{QI}(\mathcal{B})$,
and (c) the QICAS optimized basis with $D_{\mathrm{CAS}} = 8$.
The $y$-axis represents the number of orbitals with entropy larger than [Threshold $\times$ $\max_i S(\rho_i)$].
The first identifiable plateau in the low threshold regime then predicts the size of the active space.
First we focus on the HF orbitals. The earliest plateau in the HF basis predicts $D_\mathrm{CAS} = 7$, while at the optimal size $D_{\mathrm{CAS}}=8$ the HF orbitals do not exhibit a plateau.
In basis (b) where the overall entropy is minimized, the first plateau appears at $D_\mathrm{CAS}$, albeit it is not very prominent.
However, when we feed this predicted active space size into QICAS and obtain the corresponding optimized orbitals,
the plateau at $D_\mathrm{CAS}=8$ is further accentuated with a substantial extension, confirming that this is indeed the correct active space size.

\newpage
\clearpage
\onecolumngrid

\section{Tabulated Data} \label{app:data}

\begin{table*}[h]
    \centering
    \begin{tabular}{lllllll}
    \hline
    \rule{0pt}{2.6ex}
        R ($\mathrm{\AA}$) $\quad$ & HF-CASCI $\quad$ & HF-CASSCF $\quad$ & HF-QICAS-CASCI $\quad$ & Error $\quad$ & HF-MP2-CASCI $\quad$ & Error
        \\
        \hline
        \hline
        \rule{0pt}{2.6ex}
        0.90&	-75.12946759&	-75.20144209&	-75.19934023&	0.0021&	-75.18463955&	0.0168\\
        0.95&	-75.27135392&	-75.34401603&	-75.34240062&	0.0016&	-75.32828484&	0.0157\\
        1.00&	-75.37418035&	-75.44713825&	-75.44586770&	0.0013&	-75.43240495&	0.0147\\
        1.05&	-75.44681960&	-75.51972363&	-75.51869917&	0.0010&	-75.50581173&	0.0139\\
        1.10&	-75.49619999&	-75.56875202&	-75.56805195&	0.0007&	-75.55544219&	0.0133\\
        1.15&	-75.52774827&	-75.59969614&	-75.59886872&	0.0008&	-75.58676044&	0.0129\\
        1.20&	-75.54573301&	-75.61685395&	-75.61600405&	0.0008&	-75.60407385&	0.0128\\
        1.25&	-75.55352667&	-75.62360515&	-75.62280752&	0.0008&	-75.61078031&	0.0128\\
        1.30&	-75.55380393&	-75.62260941&	-75.62185183&	0.0008&	-75.60956170&	0.0130\\
        1.35&	-75.54868975&	-75.61595998&	-75.61522372&	0.0007&	-75.60253499&	0.0134\\
        1.40&	-75.53986690&	-75.60530313&	-75.60453251&	0.0008&	-75.59137073&	0.0139\\
        1.45&	-75.52865167&	-75.59193145&	-75.59115951&	0.0008&	-75.57738559&	0.0145\\
        1.50&	-75.51604788&	-75.57685721&	-75.57614399&	0.0007&	-75.56161491&	0.0152\\
        1.55&	-75.50279544&	-75.56087022&	-75.56011649&	0.0008&	-75.54486974&	0.0160\\
        1.60&	-75.48942913&	-75.54458384&	-75.54390621&	0.0007&	-75.52778131&	0.0168\\
        1.65&	-75.47498850&	-75.52847206&	-75.52755548&	0.0009&	-75.52024140&	0.0082\\
        1.70&	-75.46761485&	-75.51603912&	-75.51439989&	0.0016&	-75.51093774&	0.0051\\
        1.75&	-75.45996259&	-75.50679056&	-75.50449374&	0.0023&	-75.50138186&	0.0054\\
        1.80&	-75.45221991&	-75.49754368&	-75.49602688&	0.0015&	-75.49179947&	0.0057\\
        1.85&	-75.44449233&	-75.48847010&	-75.48691210&	0.0016&	-75.48235903&	0.0061\\
        1.90&	-75.43691689&	-75.47969780&	-75.47818189&	0.0015&	-75.47318460&	0.0065\\
        1.95&	-75.42960501&	-75.47132136&	-75.46992772&	0.0014&	-75.46436648&	0.0070\\
        2.00&	-75.42262399&	-75.46341031&	-75.46210159&	0.0013&	-75.45596985&	0.0074\\
        2.05&	-75.41605627&	-75.45601527&	-75.45499966&	0.0010&	-75.44804145&	0.0080\\
        2.10&	-75.40992543&	-75.44917221&	-75.44834985&	0.0008&	-75.44061440&	0.0086\\
        2.15&	-75.40428211&	-75.44290490&	-75.44184310&	0.0011&	-75.43371138&	0.0092\\
        2.20&	-75.39914843&	-75.43722600&	-75.43621747&	0.0010&	-75.42734653&	0.0099\\
        2.25&	-75.39451911&	-75.43213719&	-75.43156663&	0.0006&	-75.42152625&	0.0106\\
        2.30&	-75.39039472&	-75.42762900&	-75.42695443&	0.0007&	-75.41624947&	0.0114\\
        2.35&	-75.38676296&	-75.42368088&	-75.42330421&	0.0004&	-75.41150757&	0.0122\\
        2.40&	-75.38360577&	-75.42026186&	-75.41983790&	0.0004&	-75.40728441&	0.0130\\
        2.45&	-75.38087573&	-75.41733213&	-75.41692488&	0.0004&	-75.40355674&	0.0138\\
        2.50&	-75.37852375&	-75.41484545&	-75.41454286&	0.0003&	-75.40029491&	0.0146\\
        2.55&	-75.37653318&	-75.41275185&	-75.41246927&	0.0003&	-75.39746416&	0.0153\\
        2.60&	-75.37484268&	-75.41100054&	-75.41083042&	0.0002&	-75.39502619&	0.0160\\
        2.65&	-75.37341422&	-75.40954236&	-75.40939055&	0.0002&	-75.39294095&	0.0166\\
        2.70&	-75.37220493&	-75.40833169&	-75.40829325&	0.0000&	-75.39116835&	0.0172\\
        2.75&	-75.37110250&	-75.40761748&	-75.40678190&	0.0008&	-75.38966973&	0.0177\\
        2.80&	-75.37023221&	-75.40677594&	-75.40649893&	0.0003&	-75.38840902&	0.0181\\
        2.85&	-75.36949194&	-75.40607585&	-75.40587908&	0.0002&	-75.38735350&	0.0184\\
        2.90&	-75.36886008&	-75.40549195&	-75.40531755&	0.0002&	-75.38647408&	0.0188\\
        2.95&	-75.36831894&	-75.40500335&	-75.40479502&	0.0002&	-75.38574544&	0.0190\\
        3.00&	-75.36785398&	-75.40459304&	-75.40432654&	0.0003&	-75.38514585&	0.0192\\
\hline
    \end{tabular}
    \caption{HF-CASCI(8,\:8), HF-CASSCF(8,\:8), HF-QICAS-CASCI(8,\:8) energy (a.u.) and HF-MP2-CASCI(8,\:8) (a.u.) for $\mathrm{C}_2$ with cc-pVDZ basis set as functions of the internuclear distance R ($\mathrm{\AA}$). The columns directly after HF-QICAS-CASCI and HF-MP2-CASCI show the error between the corresponding CASCI energy and the HF-CASSCF energy. For QICAS the bond dimension is set at $m=100$ with 50 DMRG sweeps. For the HF-MP2-QICAS, the natural orbitals of the MP2 solution is used for the subsequent CASCI calculation. The HF-CASSCF energy at $\mathrm{R}=1.65\mathrm{\AA}$ is computed with the \textsc{Molpro}\cite{molpro1,molpro2,molpro3} package, and all other HF-CASCI/CASSCF energies with \textsc{PySCF}\cite{sun2018pyscf}.}
    \label{tab:QICAS_c2diss}
\end{table*}

\begin{table*}
    \centering
    \begin{tabular}{llll}
    \hline
    \rule{0pt}{2.6ex}
 Method($N_\mathrm{CAS}$,\:$D_\mathrm{CAS}$) $\quad\:\:\:$ & \# Micro $\quad$ & Energy (a.u.) $\quad $ &  $\langle S^2 \rangle$
        \\
        \hline
        \hline
        \rule{0pt}{3ex}
     HF-CASSCF(12,\:12) & 0 & -2098.56762004 $\quad\quad$ & 0.0000 \\
&8  &-2098.65815254 & 0.0000 \\
&16 &-2098.78440779 & 0.0000 \\
&24 &-2098.90386814 & 0.0000 \\
&32 &-2098.97374464 & 0.0000 \\
&40 &-2099.07661965 & 0.0000 \\
&48 &-2099.15989834 & 0.0000 \\
&56 &-2099.22523508 & 0.0000 \\
&66 &-2099.26654074 & 0.0000 \\
&76 &-2099.26725524 & 0.0000 \\
&80 &-2099.26725943 & 0.0000 \\
&82 &-2099.26725956 & 0.0000 \\
&83 &-2099.26725956 & 0.0000 \\

      HF-QICAS-CASSCF(12,\:12)$\quad\quad$ & 0 & -2099.21963835 & 0.0000 \rule{0pt}{3ex}\rule[0ex]{0pt}{0pt} \\
&10 &-2099.26689887 & 0.0000 \\
&16 &-2099.26725514 & 0.0000 \\
&19 &-2099.26725944 & 0.0000 \\
&21 &-2099.26725956 & 0.0000 \\
&22 &-2099.26725956 & 0.0000 \\
     HF-CASSCF(12,\:14)$\quad$ & 0 & -2098.56831064 & 0.0000 \rule{0pt}{3ex}\rule[0ex]{0pt}{0pt} \\
&8  &-2098.27411466 & 0.0231 \\
&16 &-2098.79419658 & 0.0000 \\
&24 &-2098.92256223 & 0.0000 \\
&32 &-2099.03522226 & 0.0000 \\
&40 &-2099.09255949 & 0.0000 \\
&48 &-2099.18377830 & 0.0000 \\
&56 &-2099.26174338 & 0.0000 \\
&64 &-2099.28119366 & 0.0000 \\
&72 &-2099.28981686 & 0.0000 \\
&82 &-2099.29106266 & 0.0000 \\
&92 &-2099.29109795 & 0.0000 \\
&94 &-2099.29109837 & 0.0000 \\
&95 &-2099.29109845 & 0.0000 \\
     HF-QICAS-CASSCF(12,\:14)$\quad$ & 0 & -2099.24319492 & 0.0000 \rule{0pt}{3ex}\rule[0ex]{0pt}{0pt} \\
&10 &-2099.29090519 & 0.0000 \\
&18 &-2099.29109765 & 0.0000 \\
&21 &-2099.29109853 & 0.0000 \\
&22 &-2099.29109856 & 0.0000 \\
 \hline
\end{tabular}
    \caption{Energy as a function of the number of micro iterations in CASSCF(12,12) and CASSCF(12,14) for $\mathrm{Cr_2}$ in the cc-pV5Z-DK basis set at $R=1.679\,\mathrm{\AA}$, starting from HF-orbitals and from QICAS optimized orbitals.}
    \label{tab:cr2_v5z}
\end{table*}

\clearpage

\begin{table*}
    \centering
    \begin{tabular}{llll}
    \hline
    \rule{0pt}{2.6ex}
 Method($N_\mathrm{CAS}$,\:$D_\mathrm{CAS}$) $\quad\:\:\:$ & \# Micro $\quad$ & Energy (a.u.) $\quad $ & $\langle S^2 \rangle$
        \\
        \hline
        \hline
        \rule{0pt}{3ex}
 HF-CASSCF(12,\:12) & 0 & -2098.55947438 $ \quad\quad$ & 0.0000 \rule{0pt}{3ex}\rule[0ex]{0pt}{0pt} \\
&8  &-2098.62309906 & 0.0000 \\
&16 &-2098.76339556 & 0.0000 \\
&24 &-2098.85949993 & 0.0000 \\
&32 &-2098.95413261 & 0.0000 \\
&40 &-2099.14159937 & 0.0000 \\
&48 &-2099.24342692 & 0.0000 \\
&56 &-2099.26371403 & 0.0000 \\
&66 &-2099.26724865 & 0.0000 \\
&72 &-2099.26745342 & 0.0000 \\
&76 &-2099.26745730 & 0.0000 \\
&77 &-2099.26745741 & 0.0000 \\
&78 &-2099.26745743 & 0.0000 \\
  HF-QICAS-CASSCF(12,\:12)$\quad\quad$ & 0 & -2099.11949368 & 0.0000 \rule{0pt}{3ex}\rule[0ex]{0pt}{0pt} \\
&10 &-2099.26475360 & 0.0000 \\
&18 &-2099.26726838 & 0.0000 \\
&25 &-2099.26745489 & 0.0000 \\
&28 &-2099.26745735 & 0.0000 \\
&29 &-2099.26745742 & 0.0000 \\
  HF-CASSCF(12,\:14) & 0 & -2098.55949885 & 0.0000 \rule{0pt}{3ex}\rule[0ex]{0pt}{0pt} \\
&8  &-2098.62956185 & 0.0000 \\
&16 &-2097.84547862 & 0.1102 \\
&24 &-2098.91240284 & 0.0000 \\
&32 &-2098.94968193 & 0.0000 \\
&40 &-2099.05590176 & 0.0000 \\
&48 &-2099.17568356 & 0.0000 \\
&56 &-2099.25780515 & 0.0000 \\
&64 &-2099.28634932 & 0.0000 \\
&72 &-2099.29085930 & 0.0000 \\
&82 &-2099.29122633 & 0.0000 \\
&92 &-2099.29130420 & 0.0000 \\
&96 &-2099.29130780 & 0.0000 \\
&98 &-2099.29130824 & 0.0000 \\
&99 &-2099.29130834 & 0.0000 \\
HF-QICAS-CASSCF(12,\:14) & 0 & -2099.13797545 & 0.0000 \rule{0pt}{3ex}\rule[0ex]{0pt}{0pt}  \\
&10 &-2099.29041858 & 0.0000 \\
&18 &-2099.29130378 & 0.0000 \\
&21 &-2099.29130838 & 0.0000 \\
&22 &-2099.29130846 & 0.0000 \\
\hline
    \end{tabular}
    \caption{Energy as a function of the number of micro iterations in CASSCF(12,12) and CASSCF(12,14) for $\mathrm{Cr_2}$ in the aug-cc-pV5Z-DK basis set at $R=1.679\,\mathrm{\AA}$, starting from HF-orbitals and from QICAS optimized orbitals.}
    \label{tab:cr2_av5z}
\end{table*}


\end{document}